\documentclass{easychair}
\usepackage{doc}
\usepackage{tikz}
\newtheorem{theorem}{Theorem}
\title{Mining counterexamples for wide-signature algebras with an Isabelle server
  \thanks{This project received funding from the European Research Council (ERC) under the European Union’s Horizon 2020 research and innovation program (grant agreement No. 670624). This work has also been supported by the French government, through the 3IA Côte d’Azur Investments in the Future project managed by the National Research Agency (ANR) with the reference number ANR-19-P3IA-0002. The authors would like to thank Makarius Wenzel for commenting on an early version of the Python client code.}}
\author{Wesley Fussner\inst{1} \and
Boris Shminke\inst{1}
}
\institute{Laboratoire J.A. Dieudonn\'e, CNRS, and Universit\'e C\^ote d'Azur, France\\
\email{wfussner@unice.fr}\\
\email{Boris.SHMINKE@univ-cotedazur.fr}}
\authorrunning{W. Fussner and B. Shminke}
\titlerunning{Mining counterexamples for wide-signature algebras with an Isabelle server}
\begin{document}
\maketitle
\begin{abstract}
We propose an approach for searching for counterexamples of statements about algebraic structures with a medium-sized signature using the Isabelle proof assistant in an efficient, parallel manner. We contribute a Python client Isabelle server and other scripts implementing our approach, and provide results of our computational experiments. In particular, our experiments yield counterexamples that resolve a previously open question regarding the interdependencies between distributive-like identities in residuated binars.
\end{abstract}
In partnership with automated theorem proving, finite model builders have been applied highly effectively in studies of algebraic structures (e.g., for quasigroups and loops \cite{phillips2010automated}). However, the more fundamental operations there are appearing in an algebraic language, the more expensive computations become. Most successful applications of computational tools concern semigroups, quasigroups, and other algebraic structures with only a few fundamental operations, and algebras of even slightly wider signatures pose considerable challenges (but see, e.g., \cite{SV2008,JK2019}).

This contribution offers a case study in computer-assisted counterexample construction for algebras of wider signature. Our case study concerns \emph{residuated binars} \cite{FJ2019}, each of which consists of an algebra ${\mathbf A} = (A,\wedge,\vee,\cdot,\backslash,\slash)$ such that $(A,\wedge,\vee)$ is lattice, $(A,\cdot)$ is a set with a binary operation, and for all $x,y,z\in A$,
\begin{equation}\label{eq:resid}
x\cdot y\leq z \iff y\leq x\backslash z \iff x\leq z\slash y.
\end{equation}
In this context, \cite{FJ2019} studies implications among the distributivity identities
\begin{align}
x\cdot(y\wedge z)= (x\cdot y)\wedge(x\cdot z) \label{eq:1} \\
(x\wedge y)\cdot z= (x\cdot z)\wedge (y\cdot z) \label{eq:2} \\
x\backslash (y\vee z) = (x\backslash y)\vee(x\backslash z ) \label{eq:3} \\
(x\vee y)\slash z = (x\slash z)\vee(y\slash z) \label{eq:4} \\
(x\wedge y)\backslash z = (x\backslash z)\vee(y\backslash z) \label{eq:5} \\
x\slash (y\wedge z) = (x\slash y)\vee(x\slash z) \label{eq:6}
\end{align}
in the presence of \emph{lattice distributivity}
  \begin{equation} \label{eq:dist}
    x\wedge (y\vee z)= (x\wedge y)\vee (x\wedge z).
    \end{equation}
The identities \eqref{eq:1}--\eqref{eq:6} have proven important in obtaining normal forms for terms in residuated structures. This has found applications everywhere from establishing non-trivial categorical equivalences \cite{GR2012} to obtaining decidability results for models of program execution \cite{GGMS2021}. These identities are interdependent, and \cite{FJ2019} establishes the following:
\begin{theorem}[Theorem~2.3 of \cite{FJ2019}]\label{old_theorem}
Let ${\mathbf A}$ be a residuated binar satisfying \eqref{eq:dist}. Then:
\begin{multicols}{2}
\begin{enumerate}
\itemsep0em 
\item \eqref{eq:4},\eqref{eq:5} implies \eqref{eq:3}.
\item \eqref{eq:3},\eqref{eq:6} implies \eqref{eq:4}.
\item \eqref{eq:1},\eqref{eq:4} implies \eqref{eq:6}.
\item \eqref{eq:2},\eqref{eq:3} implies \eqref{eq:5}.
\item \eqref{eq:5},\eqref{eq:1} implies \eqref{eq:2}.
\item \eqref{eq:6},\eqref{eq:2} implies \eqref{eq:1}.
\end{enumerate}
\end{multicols}
\end{theorem}
Furthermore, by exhibiting some small countermodels (of size $4$ and $5$), \cite{FJ2019} shows that the implications announced in the previous theorem completely characterize all interdependencies among \eqref{eq:1}--\eqref{eq:6} in the presence of lattice distributivity. \cite{FJ2019} mentions the case without lattice distributivity as an open question.

This contribution resolves the aforementioned question by the computer-assisted construction of finite countermodels witnessing the failure of the previous theorem in the general (i.e., non-lattice-distributive) case. We obtain:
\begin{theorem}\label{thm:main}
In general, none of the distributivity laws \eqref{eq:1}--\eqref{eq:6} follows from any combination of the others.
\end{theorem}
Thanks to its accessibility and amenability to proof simplification strategies (see \cite{K2019}), working algebraists tend to favor McCune's {\sc Prover9/Mace4} \cite{prover9-mace4} for automated work. However, {\sc Prover9/Mace4} is now regarded as somewhat out-of-date among researchers in automated deduction. Its model search capability has been gradually surpassed by a series of improvements in the field. The Paradox \cite{Claessen2007NewTT} system introduced so-called static symmetry reduction, a technique reducing the number of isomorphic models (see \cite{Baumgartner07computingfinite} for {\sc Mace4} and Paradox comparison). Later, Kodkod (see \cite{Torlak07kodkod:a} for realization details and comparison with Paradox) brought sparse representation of binary relations and even more symmetry-breaking schemes to the process of translating a model-search task into a SAT problem.

We obtain Theorem~\ref{thm:main} with the help of Nitpick, a highly efficient tool for the construction of finite counterexamples packaged with the Isabelle proof assistant \cite{isabelle-system}. Nitpick serves as a translator from Isabelle language to Kodkod, which currently relies on Jingeling (\cite{Biere-SAT-Competition-2017-solvers}, the winner of SAT 2020 competition). Our work takes advantage of the fact that the Isabelle server implementation can run Nitpick tasks in parallel, yielding an environment for countermodel search with large computational advantages. Nevertheless, we observed that model search for residuated binars in cardinalities higher than 14 did not finish even after a week of running an Isabelle server. For {\sc Mace4} the boundary was lower, and even models of size larger than 5 became intractable.

To our best knowledge, there was no established way to communicate with the Isabelle server from Python. Thus we created a Python client to the Isabelle server \cite{isabelle-client} and a parser of Isabelle server logs (from TCP-communicated JSON parcels to \LaTeX\ code for Cayley tables and Hasse diagrams). We conducted our computational experiments yielding Theorem~\ref{thm:main} on three Linux machines, the largest having 180 CPU cores (\textsc{Intel}\textsuperscript{\textregistered} \textsc{Xeon}\textsuperscript{\textregistered} Gold 6254 3.10GHz) and 832 GB of RAM, totaling to about two weeks of wall-clock time. The conclusion of these experiments resulted in mined counterexamples (the largest having the underlying set of $10$ elements) supporting the proof of Theorem~\ref{thm:main}. These counterexamples and the code for getting them are available as supplementary material for this paper.\footnote{\url{https://github.com/inpefess/residuated-binars}}

\label{sect:bib}
\bibliographystyle{plain}
\bibliography{aitp2021_2021.08.13}

\begin{thebibliography}{10}

\bibitem{Baumgartner07computingfinite}
P.~Baumgartner, A.~Fuchs, H.~{de Nivelle}, and C.~Tinelli.
\newblock Computing finite models by reduction to function-free clause logic.
\newblock {\em J. Appl. Log.}, 7(1):58--74, 2009.

\bibitem{Biere-SAT-Competition-2017-solvers}
A.~Biere.
\newblock {CaDiCaL, Lingeling, Plingeling, Treengeling, YalSAT Entering the SAT
  Competition 2017}.
\newblock In T.~Balyo, M.~Heule, and M.~J{\"a}rvisalo, editors, {\em Proc.~of
  {SAT Competition} 2017 -- Solver and Benchmark Descriptions}, volume B-2017-1
  of {\em Department of Computer Science Series of Publications B}, pages
  14--15. University of Helsinki, 2017.

\bibitem{Claessen2007NewTT}
K.~Claessen and N.~S{\"o}rensson.
\newblock New techniques that improve {MACE}-style finite model finding.
\newblock In {\em Proceedings of the CADE-19 Workshop: Model
  Computation-Principles, Algorithms, Applications}, pages 11--27. Citeseer,
  2003.

\bibitem{FJ2019}
W.~Fussner and P.~Jipsen.
\newblock Distributive laws in residuated binars.
\newblock {\em Algebra Universalis}, 80.54, 2019.

\bibitem{GR2012}
N.~Galatos and J.G. Raftery.
\newblock A category equivalence for odd {S}ugihara monoids and its
  applications.
\newblock {\em J. Pure Appl. Algebra}, 216:2177--2192, 2012.

\bibitem{JK2019}
P.~Jipsen and M.~Kinyon.
\newblock Nonassociative right hoops.
\newblock {\em Algebra Universalis}, 80.47, 2019.

\bibitem{K2019}
M.~Kinyon.
\newblock Proof simplification and automated theorem proving.
\newblock {\em Philos. Trans. Roy. Soc. A}, 377.20180034, 2019.

\bibitem{prover9-mace4}
W.~McCune.
\newblock {P}rover9 and {M}ace4.
\newblock \verb|http://www.cs.unm.edu/~mccune/prover9/|, 2005--2010.

\bibitem{phillips2010automated}
J.D. Phillips and D.~Stanovsk{\'y}.
\newblock Automated theorem proving in quasigroup and loop theory.
\newblock {\em {AI} Communications}, 23:267--283, 2010.

\bibitem{isabelle-client}
B.~Shminke.
\newblock Python client for isabelle server.
\newblock \url{https://pypi.org/project/isabelle-client/}.

\bibitem{SV2008}
M.~Spinks and R.~Veroff.
\newblock Constructive logic with strong negation is a substructural logic {I}.
\newblock {\em Studia Logica}, 88:325--348, 2008.

\bibitem{Torlak07kodkod:a}
E.~Torlak and D.~Jackson.
\newblock Kodkod: A relational model finder.
\newblock In {\em In Tools and Algorithms for Construction and Analysis of
  Systems (TACAS)}, pages 632--647. Wiley, 2007.

\bibitem{GGMS2021}
S.~van Gool, A.~Guatto, G.~Metcalfe, and S.~Santschi.
\newblock Time warps, from algebra to algorithms.
\newblock 2021.
\newblock Preprint. Available at arXiv:2106.06205.

\bibitem{isabelle-system}
M.~Wenzel.
\newblock {\em The {Isabelle} System Manual}.
\newblock \url{https://isabelle.in.tum.de/doc/system.pdf}.

\end{thebibliography}
\end{document}